\documentclass[subscriptcorrection,upint,varvw,nolists]{asmejour}

\hypersetup{%
	pdfauthor={Arunav Choudhury, R. Ganesh},                       		   	% <=== change to YOUR name[s]!
	pdftitle={Self-excited oscillations in multi-degree-of-freedom systems subjected to discontinuous forcing},                  	% <=== change to YOUR pdf file title
	pdfkeywords={limit cycles, non-smooth dynamics, method of averaging, bifurcation, piecewise-linear systems},% <=== change to YOUR pdf keywords
	pdfsubject = {Nonlinear Dynamics},			% <=== change to YOUR subject
}

\begin{document}

% Change to your author name[s] and addresses, in the desired order of authors.
% First name, middle initial, last name
% Use title case (upper and lower case letters)
% Note usage below for corresponding author.

\newcommand{\IITBaffil}{%
Department of Mechanical Engineering,\\
Indian Institute of Technology Bombay,\\
Powai, Mumbai, Maharashtra, India}

\SetAuthorBlock{Arunav Choudhury}{%
\IITBaffil\\ email: 214106002@iitb.ac.in}

\SetAuthorBlock{R. Ganesh\CorrespondingAuthor}{%
\IITBaffil\\ email: ganeshr@iitb.ac.in}

%%% Change to your paper title. Can insert line breaks if you wish (otherwise breaks are selected automatically).
\title{Self-excited oscillations in multi-degree-of-freedom systems subjected to discontinuous forcing}

%%% Change these to your keywords.  Keywords are automatically printed at the end of the abstract.
%%% This command must come BEFORE the end of the abstract.
%%% If you don't want keywords, omit the \keyword{..} command.
   
%% Abstract should be no more than 250 words
\begin{abstract}
This study investigates the existence and stability of limit cycles resulting from self-excited oscillations in linear multi-degree-of-freedom systems subjected to discontinuous, state-dependent forcing. Using the method of averaging and slow-flow phase-plane analysis, analytical expressions are derived for the amplitudes and stability boundaries of limit cycles in a two-degree-of-freedom system. The analysis demonstrates that stable limit cycles may exist in all natural modes, with the steady-state response governed by initial conditions in regimes of multistability. A central contribution of this work is the identification and analytical characterization of the stability-axis-flipping (SAF) bifurcation, which serves as the governing mechanism for the exchange of stability between modes. The framework is then systematically extended to systems with higher degrees of freedom, confirming that the SAF bifurcation remains a universal feature, even under varying feedback configurations. The steady-state dynamics, summarized through stability maps and validated by numerical simulations, delineate the existence and stability regions of modal limit cycles as functions of key system parameters. These results provide efficient criteria for guiding optimization studies to mitigate or generate limit cycles at targeted frequencies in flexible mechanical structures.
\end{abstract}

\date{}%% You can modify this information as desired. 
							%% Putting \date{} will suppress any date.  
							%% If this command is omitted, date defaults to \today
							%% This command must come somewhere before \maketitle

\maketitle %% This command creates the author/title/abstract block. Essential!

%%%%%%%%%%%%%%%%%%%%%%%%%%%%%%%%%%%%%%%%%%%%%%%%%%%%%%%%%%%%%%%%%%%%%%%%%%%%%%%%%%%%%%%%%%%%%%%%%%%%%%%

\section{Introduction}
\label{Intro}
Mechanical structures possess multiple modes of oscillation, and their dynamic response is fundamentally governed by the nature of the applied forcing. Discontinuous or non-smooth nonlinearities arising from mechanisms such as dry friction, mechanical impacts, or switching elements act as state-dependent feedback that can introduce abrupt qualitative changes in system dynamics. For example, structural pounding between the slabs of adjacent buildings due to out-of-phase vibrations during earthquakes leads to plastic deformations and severe structural damage~\cite{Langlade2021}. Similarly, aeroelastic components in gas turbines and aircraft control surfaces may experience flutter-induced limit cycles that undergo sudden, large-amplitude transitions driven by intrinsic non-smooth nonlinear effects~\cite{kumar2021dynamics,vasconcellos_aeroelastic_2022,petrov_analysis_2012}.

From a structural design perspective, the emergence of stable limit cycles is of practical concern. Such oscillations persist in the absence of external periodic excitation, and may lead to accelerated wear, performance degradation, and cumulative fatigue damage~\cite{nayfeh2024nonlinear}. The associated risk can become more pronounced when the dominant oscillation shifts from the fundamental to a higher mode, wherein control or mitigation strategies designed for the primary mode may become less effective~\cite{Ramakrishnan2019}. Consequently, limit cycles induced by discontinuous or non-smooth feedback mechanisms present challenges in engineering design and motivate careful analysis of their existence and stability.

Existing literature extensively documents the fundamental mechanisms of self-excited oscillations in single-degree-of-freedom (SDOF) systems subjected to discontinuous forcing~\cite{filippov1960differential,Andronov1966,bernardo2008piecewise}. Within this framework, analytical approximations characterize stick--slip oscillations and capture the interaction between stick and slip phases in the steady-state response~\cite{JUELTHOMSEN2003389}. Further, non-stiff switching formulations that bridge theory and numerical implementation enable efficient computation of limit cycles in stick--slip systems~\cite{Leine1998}. Subsequent studies extend these deterministic approaches to stochastic settings by examining base-driven dry-friction oscillators under random excitation and developing statistical descriptions of stick--slip phase sequences and durations~\cite{Lima2017}. More recent work on piecewise-smooth systems shows that switching across constraint boundaries can generate or modify limit cycle oscillations through non-smooth bifurcation mechanisms~\cite{Cao2024}.

In contrast, relatively few studies investigate the emergence and stability of limit cycles across different modes in non-smooth, multi-degree-of-freedom (MDOF) systems, primarily due to the analytical complexity associated with multi-dimensional switching boundaries. Existing research typically focuses on specific phenomena rather than general stability mechanisms. For example, multi-harmonic frequency-domain approaches compute steady-state responses of MDOF systems with Coulomb friction~\cite{Pierre1985}, while hybrid computational frameworks enable efficient evaluation of full-order dynamics in piecewise-linear friction systems without modal reduction~\cite{Shahhosseini2023}. However, these studies predominantly consider harmonically forced systems in which friction drives the nonlinear response. Other investigations examine systems with structural discontinuities and demonstrate the occurrence of limit cycles with varying amplitudes~\cite{Rehan2025}. Despite these advances, a general analytical framework that explains modal interaction and the exchange of stability between competing modal limit cycles remains unavailable.

To address these limitations, the present work develops analytically tractable non-smooth MDOF models that enable rigorous identification of the conditions under which stable limit cycles arise in higher modes. The approach retains sufficient mathematical structure to support closed-form analysis and qualitative interpretation while capturing the essential dynamics introduced by discontinuous forcing within an otherwise linear MDOF framework.

In Section \ref{Sign}, a relay-type feedback forcing characterized by constant magnitude with velocity-dependent switching \cite{aguilar2015self} is applied to a two-degree-of-freedom system. The analysis reveals that modal limit cycles exchange stability through a structured sequence of qualitative changes, defined here as a stability-axis-flipping (SAF) bifurcation. This remains the governing mechanism even as the number of degrees of freedom is increased, demonstrating the robustness of the bifurcation in higher-order systems.

Section \ref{section_3} introduces a modified feedback mechanism based on discontinuous negative damping. The resulting dynamics exhibit selective modal destabilization, wherein the stability exchange between different modes is governed by the ratio of damping parameters. The SAF bifurcation again characterizes these transitions, illustrating the generality of the mechanism across different forcing types.

The results from both analyses are integrated into stability maps that provide predictive criteria for engineering applications. These maps serve as a guide for suppressing undesirable higher-mode oscillations~\cite{Dwars2015}, or for inducing stable periodic responses in applications ranging from nonlinear control~\cite{aguilar2009generating} to vibration-based energy harvesting~\cite{WANG2018339}.

\section{Mathematical Model}
\label{Sign}
\begin{figure*}[!htb]
	\centering
	\includegraphics[width=.9\linewidth]{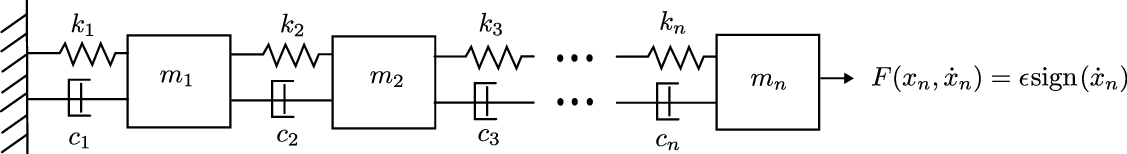}
	\caption{Schematic representation of a one-dimensional spring-mass model subjected to an implicit terminal forcing}
	\label{fig:1}
\end{figure*}
Consider a one-dimensional structure that is discretized as a series of masses connected by linear springs and dampers, as shown in Fig.~\ref{fig:1}. The number of degrees of freedom is arbitrary ($n$), and the equations of motion for this system are
\begin{equation}
	\mathbf{M} \ddot{\mathbf{x}} + \mathbf{C} \dot{\mathbf{x}} + \mathbf{K} \mathbf{x} = \mathbf{f}(t),
	\label{eq:1}
\end{equation}
where $\mathbf{x} = [x_1, x_2, x_3, \cdots, x_n]^T$ represents the displacement of each mass from equilibrium, and $\mathbf{M}$, $\mathbf{K}$ are the mass and stiffness matrices, respectively, given by
\begin{equation*}
\begin{aligned}
\mathbf{M} &=
\begin{bmatrix}
m_{1} & 0 & \cdots & 0 \\
0 & m_{2} & \cdots & 0 \\
\vdots & \vdots & \ddots & \vdots \\
0 & 0 & \cdots & m_{n}
\end{bmatrix}, \\[6pt]
\mathbf{K} &=
\begin{bmatrix}
k_{1}+k_{2} & -k_{2} & 0 & \cdots & 0 \\
-k_{2} & k_{2}+k_{3} & -k_{3} & \cdots & 0 \\
0 & -k_{3} & \ddots & \ddots & \vdots \\
\vdots & \vdots & \ddots & k_{n-1}+k_{n} & -k_{n} \\
0 & 0 & \cdots & -k_{n} & k_{n}
\end{bmatrix}.
\end{aligned}
\end{equation*}
The damping matrix is assumed to satisfy proportional damping $(\mathbf{C} = \beta \, \mathbf{K})$, and the system is subjected to an implicit, discontinuous forcing applied only to the end mass, given by
\begin{equation*}   
	\mathbf{f}(\mathbf{x},\dot{\mathbf{x}},t) = [0,0,\cdots,0,\varepsilon \ \text{sign}(\dot{x}_{n})]^T,
\end{equation*}
where $\varepsilon > 0$ is a small parameter that represents the strength of the discontinuous nonlinearity. If $\varepsilon$ were negative, the force would be identical to that generated by rate-independent frictional contact on the tip mass, which would cause oscillations to decay in the absence of an external energy source. In contrast, the adopted forcing is opposite in nature to Coulomb friction and is inherently destabilizing; any small perturbation of the trivial equilibrium leads to a self-excited limit cycle in the system. Similar forcing functions have been employed to analyze mechanical systems with relay feedback~\cite{aguilar2015self}.

Since the forcing is assumed to be weak, the method of averaging can be employed to analyze the slow evolution of the modal amplitudes. First, the equations of motion (Eq.~\ref{eq:1}) are transformed into modal coordinates by introducing $\mathbf{x} = \boldsymbol{\Phi}\mathbf{y}$, where $\boldsymbol{\Phi}$ is the mass-normalized modal matrix whose columns represent the eigenvectors of the natural modes of the undamped, linear system. Then, premultiplying by $\boldsymbol{\Phi}^{T}$ decouples the equations in the modal coordinates; for periodic solutions defined by $y_{i}(t)=A_{i}(t)\cos(\theta_i(t))$ $(i=1,2,\dots,n)$, the amplitudes $A_i$ are obtained using the method of averaging as
\begin{equation}
\dot{A}_{i} = -\frac{1}{2}\beta \omega_{i}^{2} A_{i} + T_{i}, \quad i=1,\dots,n, 
\label{eq:2}
\end{equation}
where
\begin{align*}
T_{i} &= \frac{\epsilon \Phi_{ni}}{(2\pi)^{n}\omega_{i}}
\int_{0}^{2\pi}\!\cdots\!\int_{0}^{2\pi}
\tilde{T}_{i}~\mathrm{d}\theta_{1}\cdots \mathrm{d}\theta_{n}, \\[4pt]
\tilde{T}_{i} & = \mathrm{sign}\!\left(\sum_{j=1}^{n}
\Phi_{nj}\,\omega_{j}\,A_{j}\,\sin\theta_{j}\right)
\sin\theta_{i}\,.
\end{align*}
Obtaining a closed-form solution using the averaging method becomes intractable for systems with an arbitrary number of degrees of freedom, primarily due to the complexity of evaluating $n$-dimensional integrals involving discontinuities. To address this limitation, we first consider a single-degree-of-freedom (SDOF) system as the baseline. We then systematically scale the model complexity by adding degrees of freedom, allowing for a progressive investigation of the resulting dynamical behavior.
\subsection{Single-Degree-of-Freedom (SDOF) System}
The equation of motion for the SDOF system is given by
\[ \ddot{x} + \beta \omega^{2} \dot{x} + \omega^{2} x = \varepsilon \, \text{sign}(\dot{x}), \]
and the slow-flow equation for the amplitude envelope can be obtained as
\[ \dot{A} = f(A) = -\dfrac{\beta \omega^{2} A}{2} + \dfrac{2 \varepsilon}{\pi \omega}. \]
The critical point for the amplitude $A$ and its Jacobian are given by
\[ A^{*} = \dfrac{4 \varepsilon}{\pi \beta \omega^{3}}, \;\; \left. \dfrac{d \dot{A}}{dA} \right|_{A=A^*} = -\dfrac{\beta \omega^{2}}{2}. \]
Since the critical point in the slow-flow phase plane corresponds to a limit cycle in the physical system, the limit cycle for the SDOF system is always stable. Consequently, any small perturbation of the trivial equilibrium $(x = \dot{x} = 0)$ leads to a limit cycle oscillation, with the steady-state amplitude directly proportional to the magnitude of the discontinuous forcing, and inversely proportional to the product of the damping factor and the cube of the natural frequency.
\subsection{Two-Degree-of-Freedom (2-DOF) system}
The equations of motion in modal coordinates for the 2-DOF system are 
\begin{equation*}
\begin{aligned}
\mathbf{I}\,\ddot{\mathbf{y}}
+ \mathbf{\beta \tilde{K}}\,\dot{\mathbf{y}}
+ \mathbf{\tilde{K}}\,\mathbf{y}
= \mathbf{f}(\dot{\mathbf{y}}),
\end{aligned}
\end{equation*}
where
\begin{align*}
\mathbf{I} &=
\begin{bmatrix}
1 & 0\\
0 & 1
\end{bmatrix}, \,
\mathbf{\tilde{K}} =
\begin{bmatrix}
\omega_{1}^{2} & 0\\
0 & \omega_{2}^{2}
\end{bmatrix}, \\[4pt]
\mathbf{y} &=
\begin{Bmatrix}
y_{1}\\
y_{2}
\end{Bmatrix}, \,
\mathbf{f}(\dot{\mathbf{y}}) =
\boldsymbol{\Phi}^{T}
\begin{Bmatrix}
0\\
\varepsilon\,\mathrm{sign}(\Phi_{21}\dot{y}_{1}+\Phi_{22}\dot{y}_{2})
\end{Bmatrix},
\end{align*}
and the dynamics in the slow-flow plane can be obtained as  
\begin{align}
\dot{A}_{1} &=
\begin{cases}
-\dfrac{\beta \omega_{1}^2 A_{1}}{2} + \dfrac{4\varepsilon \Phi_{21}}{\omega_{1}\pi^{2}} \mathcal{E}(\delta^{2}), & \text{for } \delta < 1,\\[10pt]
-\dfrac{\beta \omega_{1}^2 A_{1}}{2} + \dfrac{4\varepsilon \Phi_{21}}{\omega_{1}\pi^{2}} \delta\,f_{1}, & \text{for } \delta > 1,
\end{cases} \label{eq:3} \\[8pt]
\dot{A}_{2} &=
\begin{cases}
-\dfrac{\beta \omega_{2}^2 A_{2}}{2} + \dfrac{4\varepsilon \Phi_{22}}{\omega_{2}\pi^{2}}\dfrac{f_{2}}{\delta}, & \text{for } \delta < 1,\\[10pt]
-\dfrac{\beta \omega_{2}^2 A_{2}}{2} + \dfrac{4\varepsilon \Phi_{22}}{\omega_{2}\pi^{2}}\mathcal{E}(\delta^{-2}), & \text{for } \delta > 1,
\end{cases} \label{eq:4}
\end{align}
where
\begin{align*}
f_{1} &= \left[\mathcal{E}(\delta^{-2}) - f_{3}\mathcal{K}(\delta^{-2})\right], \, f_{3} = (1-\delta^{-2}),  \\
f_{2} &= \left[\mathcal{E}(\delta^{2}) - f_{4}\mathcal{K}(\delta^{2})\right], \, f_{4} = (1-\delta^{2}).
\end{align*}
$\mathcal{K}$ and $\mathcal{E}$ are complete elliptic integrals of the first and second kind, respectively, and the parameter $\delta$ represents the ratio of modal contributions, given by
\[ \delta=\dfrac{\Phi_{22}A_{2}\omega_{2}}{\Phi_{21}A_{1}\omega_{1}}.\]
To determine the critical points for the slow-flow equations, Eqs.~(\ref{eq:3}) and (\ref{eq:4}) can be simplified as
\begin{equation}
F(\delta) = \begin{cases}
            f_{5}\mathcal{E}(\delta^2)+f_{4}\mathcal{K}(\delta^2), & \text{if } \delta < 1,\\
            f_{6}\mathcal{E}(\delta^{-2})-f_{3}\mathcal{K}(\delta^{-2}), & \text{if } \delta > 1,
            \end{cases}
          =  0,
\label{eq:5}
\end{equation}
where
\begin{align*}
f_{5} &= \big(q^{-1}\delta^2-1\big),\,f_{6} = \big(1-q\delta^{-2}\big). 
\end{align*}
$q$ is a system parameter that is a function of the natural frequencies and mode shapes, expressed as
\[q = \left( \dfrac{\Phi_{22}\omega_{1}}{\Phi_{21}\omega_{2}} \right)^2.\] 

\begin{figure*}[!htb]
	\centering
	\includegraphics[width=.9\linewidth,keepaspectratio]{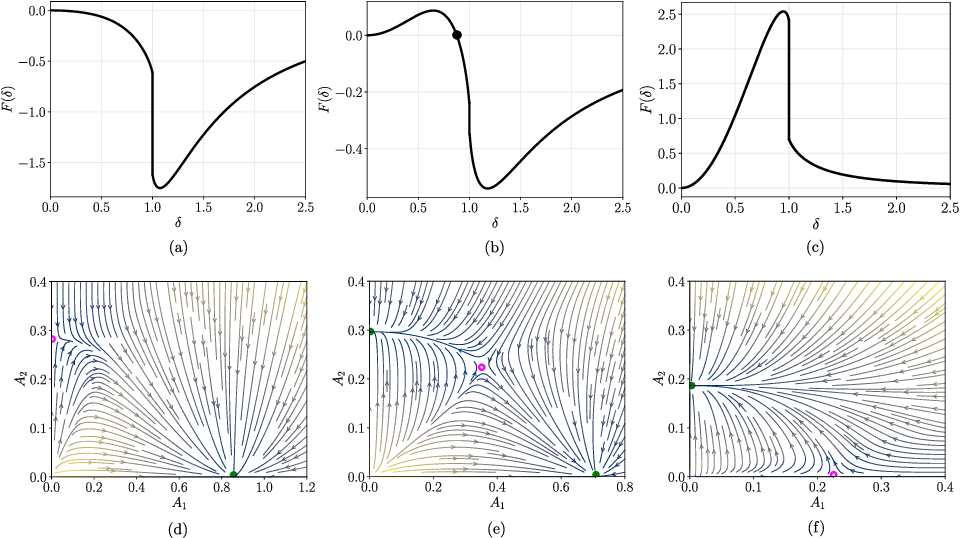}	
	\caption{Functional plots of $F(\delta)$ (Eq.~\ref{eq:5}, top row) and slow-flow phase portraits (second row) for varying values of $q$: (a, d) $q = 0.39$; (b, e) $q = 0.75$; and (c, f) $q = 3.4$. Note that $F(\delta)$ is discontinuous at $\delta=1$}
	\label{fig:2}
\end{figure*}

Figure~\ref{fig:2} shows the plot of Eq.~(\ref{eq:5}) and the dynamics in the slow-flow phase plane for different values of the system parameter $q$. For all values of $q$, two solutions always exist, corresponding to the limiting cases $\delta = 0$ and $\delta \to \infty$. The solution for $\delta = 0$ represents the critical point $(A_{1}^*, 0)$, while $\delta \to \infty$ corresponds to $(0, A_{2}^*)$. This conclusion is further supported by the phase portraits shown in Fig.~\ref{fig:2}. These critical points, which represent limit cycles in the decoupled modes, are always present in this 2-DOF system. The amplitudes of these limit cycles and the eigenvalues of the Jacobian matrix at the corresponding critical points are given by
\begin{align}
	&\mathbf{(A_{1}^{*},0)}: \, A_{1}^* = \dfrac{4\epsilon \Phi_{21}}{\beta \pi \omega_{1}^3},\; \lambda_{1} = -\dfrac{\beta \omega_{1}^2}{2},\; \lambda_{2} =  -\dfrac{\beta \omega_{2}^2}{4}\left(2 - q \right), \label{Eq:6} \\[4pt]
	&\mathbf{(0,A_{2}^{*})}:\, A_{2}^* = \dfrac{4\epsilon \Phi_{22}}{\beta \pi \omega_{2}^3},\; \lambda_{1} = -\dfrac{\beta \omega_{1}^2}{4}\left(2 - \dfrac{1}{q} \right),\; \lambda_{2} = -\dfrac{\beta \omega_{2}^2}{2}. \label{Eq:7} 
\end{align}
From Eqs.~(\ref{Eq:6}) and (\ref{Eq:7}), the critical point $(A_{1}^{*},0)$ in the slow-flow phase plane is a stable node if $q<2$, and a saddle point if $q>2$. Similarly, the critical point $(0,A_{2}^{*})$ is a stable node if $q>0.5$, and a saddle point otherwise. Consequently, both the critical points are stable nodes when
\begin{equation}
	0.5 < q < 2. \label{Eq:8}
\end{equation} 
Since the two critical points cannot simultaneously be saddle points, at least one of the single-mode limit cycle oscillations remains stable in this system. When both critical points are stable ($0.5 < q < 2$), an additional critical point must exist that acts as a saddle point, separating the basins of attraction of the two stable nodes. This is confirmed in Fig.~\ref{fig:2}(b), where an additional zero crossing of the function $F(\delta)$ is observed, corresponding to a solution of the form $(\bar{A}_{1}, \bar{A}_{2})$ in modal coordinates. Furthermore, this solution exists only when Eq.~(\ref{Eq:8}) is satisfied, implying that the associated critical point is always a saddle point. Since saddle points in the slow-flow correspond to unstable limit cycles in the physical coordinates, the mixed-mode solution in the slow-flow plane, which represents a quasi-periodic oscillation in the physical system, cannot be realized. In summary, the overall dynamics of the 2-DOF system can be completely characterized in the slow-flow phase plane, and the qualitative phase portrait as a function of the parameter $q$ is shown in Fig.~\ref{fig:2}. 

\begin{figure*}[!htb]
	\centering
	\includegraphics[width=.9\linewidth]{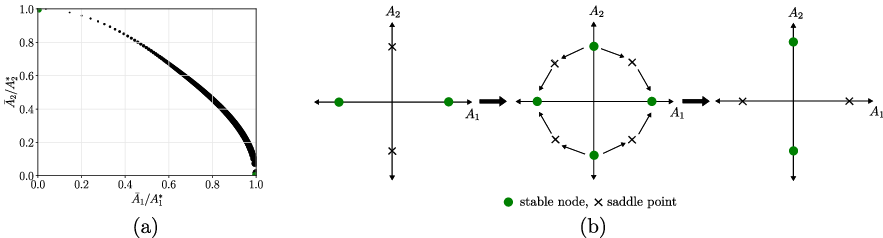}
	\caption{(a) Locus of the mixed-mode critical point $(\bar{A}_{1}, \bar{A}_{2})$ in the first quadrant of the slow-flow phase plane. (b) Bifurcation sequence for $q \in [0.5, 2]$, illustrating the SAF bifurcation}
	\label{fig:3}
\end{figure*}

To determine the root locus depicting the evolution of the mixed-mode saddle point $(\bar{A}_{1}, \bar{A}_{2})$, the numerical values of this critical point are computed using nonlinear root-solving techniques, and its variation with respect to $q$ is shown in Fig.~\ref{fig:3}(a). The mixed-mode solution $(\bar{A}_{1}, \bar{A}_{2})$ originates from the critical point $(0, A_{2}^*)$ at $q = 0.5$ and progressively approaches the critical point $(A_{1}^*, 0)$ as $q$ increases. At $q = 2$, the mixed-mode solution coalesces with this critical point and subsequently disappears for larger values of $q$. Owing to the symmetry of the system, this behavior is mirrored in all four quadrants of the slow-flow phase plane. Consequently, as $q$ crosses $0.5$, the unstable critical point $(0,A_{2}^*)$ undergoes a subcritical pitchfork bifurcation, splitting into a stable node $(0, A_{2}^*)$, and two saddle points $(\pm \bar{A}_{1}, \bar{A}_{2})$. Similarly, $(0,-A_{2}^*)$ splits into $(0,-A_{2}^*)$ and $(\pm \bar{A}_{1}, -\bar{A}_{2})$. These four saddle points $(\pm \bar{A}_{1},\pm \bar{A}_{2})$ eventually merge with the stable critical points $(\pm A_{1}^*,0)$ as $q$ approaches $2$, in a sequence constituting another subcritical pitchfork bifurcation. We use the term \emph{stability-axis-flipping (``SAF'') bifurcation} to denote the parameter-driven exchange of stability between the two axial attractors, mediated by the mixed-mode saddle that successively collides with them (subcritical pitchforks at $q=0.5$ and $q=2$), thereby flipping the stable axis of attraction in the $(A_1,A_2)$ plane. This global qualitative change, or  ``SAF'' bifurcation, is illustrated in Fig.~\ref{fig:3}(b). 

In summary, as the number of degrees of freedom increases to two, the system exhibits at least two limit cycles corresponding to the single-mode oscillations, which may be stable or unstable depending on the system parameters. When limit cycle oscillations in both modes are stable, the system is bistable, and the response converges to one of these fundamental modes depending on the initial conditions. Finally, the quasi-periodic oscillation corresponding to the mixed-mode solution cannot be realized, as the associated critical point is inherently unstable.
\subsection{Three-Degree-of-Freedom (3-DOF) system}
When the number of degrees of freedom in the system increases to three, to the best of our knowledge, no closed-form analytical solutions are available for the averaged equations. However, based on the analysis of the 2-DOF system, the slow-flow dynamics can still be investigated for specific solutions of the 3-DOF system.
\subsubsection{\textbf{Critical points on the axes}}
\label{cr_pts_axes}
Assuming two of the amplitude values to be zero in the slow-flow equations, the averaged equations for each modal coordinate can be obtained (Eq.~(\ref{eq:2}) for $n=3$). In the three-dimensional slow-flow phase space, there are three critical points on the coordinate axes: $(A_{1}^*,0,0)$, $(0,A_{2}^*,0)$ and $(0,0,A_{3}^*)$. The values of the critical points and the corresponding eigenvalues obtained by evaluating the Jacobian are:
\begin{align*}
&\mathbf{(A_{1}^{*},0,0)}:\;
A_1^{*} = \frac{4\varepsilon\Phi_{31}}{\beta\pi\omega_{1}^3},\;
\lambda_{1} = -\frac{\beta\omega_{1}^2}{2} \\
&\lambda_{2} = -\frac{\beta\omega_{2}^2}{4}(2-q_{1}),\
\lambda_{3} = -\frac{\beta\omega_{3}^2}{4}(2-q_{2}) \\[6pt]
&\mathbf{(0,A_{2}^{*},0)}:\;
A_2^{*} = \frac{4\varepsilon\Phi_{32}}{\beta\pi\omega_{2}^3},\;
\lambda_{1} = -\frac{\beta\omega_{1}^2}{4q_{1}}(2q_{1}-1) \\
&\lambda_{2} = -\frac{\beta\omega_{2}^2}{2},\;
\lambda_{3} = -\frac{\beta\omega_{3}^2}{4q_{1}}(2q_{1}-q_{2}) \\[6pt]
&\mathbf{(0,0,A_{3}^{*})}:\;
A_3^{*} = \frac{4\varepsilon\Phi_{33}}{\beta\pi\omega_{3}^3},\;
\lambda_{1} = -\frac{\beta\omega_{1}^2}{4q_{2}}(2q_{2}-1) \\
&\lambda_{2} = -\frac{\beta\omega_{2}^2}{4q_{2}}(2q_{2}-q_{1}),\;
\lambda_{3} = -\frac{\beta\omega_{3}^2}{2}.
\end{align*}
where $q_1$ and $q_2$ are system parameters given by
\[q_{1} = \left(\frac{\Phi_{32}\omega_{1}}{\Phi_{31}\omega_{2}}\right)^2,\quad q_{2} = \left(\frac{\Phi_{33}\omega_{1}}{\Phi_{31}\omega_{3}}\right)^2.\]
\begin{table}[!ht]
\caption{Stability criteria for the axial critical points of the 3-DOF slow-flow equations. These points correspond to single-mode limit cycles in the physical coordinates.}
\label{tab1}
\centering
\renewcommand{\arraystretch}{1.3}
\begin{tabular}{@{}l l @{\quad} c @{\quad} l @{}} % Defined 4 columns
\toprule
\textbf{Mode} & \multicolumn{3}{l}{\textbf{Criteria for stability}} \\ % Header 'b' spans the 3 math columns
\midrule
1$^{\text{st}}$ mode $(A_{1}^{*},0,0)$ & $q_{1} < 2$ & and & $q_{2} < 2$ \\
\addlinespace[6pt]
2$^{\text{nd}}$ mode $(0,A_{2}^{*},0)$ & $q_{1} > 0.5$ & and & $\displaystyle \frac{q_{2}}{q_{1}} < 2$ \\
\addlinespace[6pt]
3$^{\text{rd}}$ mode $(0,0,A_{3}^{*})$ & $q_{2} > 0.5$ & and & $\displaystyle \frac{q_{2}}{q_{1}} > 0.5$ \\
\bottomrule
\end{tabular}
\end{table}

The stability criterion for these critical points are summarized in Table~\ref{tab1}. Based on these criteria, at least one critical point will always be a stable node for all values of $q_1$ and $q_2$. In other words, all the critical points cannot be saddle points simultaneously. Furthermore, there are three broad categories of stability which determine the various limit cycle oscillations that are possible in the physical system (coordinates).

The first category occurs when only one critical point is stable; in this case, the limit cycle oscillation in the physical system will occur at the corresponding natural frequency. We denote this category as SUU/USU/UUS, where each letter indicates the stability (Stable/Unstable) of the limit cycle oscillation in modes 1, 2, and 3, respectively. 

The second category corresponds to a bistable system, with limit cycle oscillations in two of the three natural modes, classified as SSU/USS/SUS. In this case, the amplitude and frequency of the limit cycle oscillations will depend on the initial conditions. 

\begin{figure}[!htb]
	\centering
	\includegraphics[width=\linewidth]{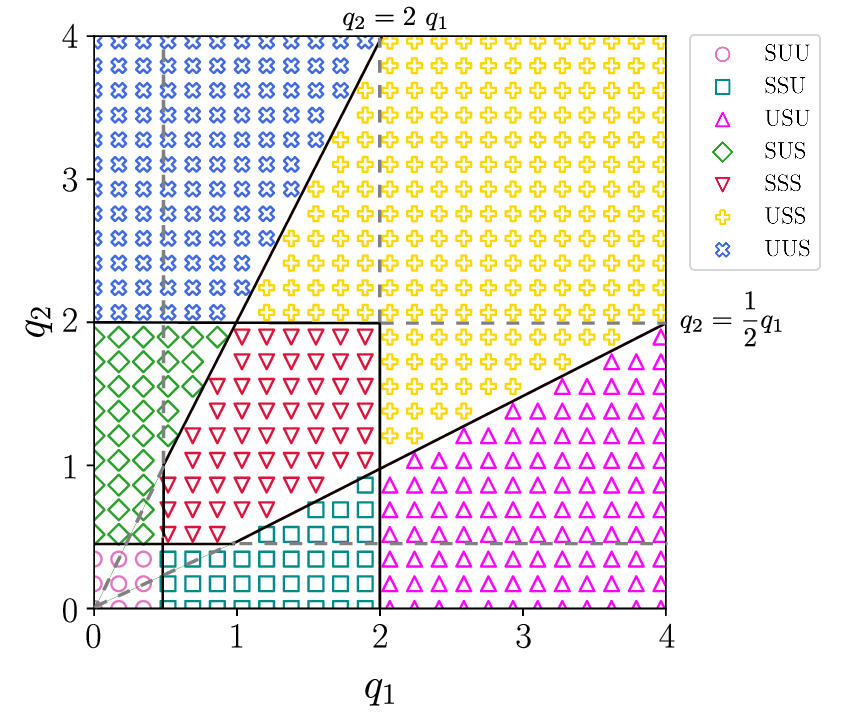}
    \caption{Stability characteristics of the three single-mode limit cycles in the $q_1$--$q_2$ parameter plane. S and U denote stable and unstable limit cycles, respectively}
	\label{fig:4}
\end{figure}

The final category corresponds to a tristable system, with limit cycle oscillations possible in all the three modes (SSS). A graphical representation of these categories in the codimension-2 parameter space is shown in Fig.~\ref{fig:4}. 
\subsubsection{\textbf{Critical points in the plane}}
We now consider critical points corresponding to mixed-mode solutions involving two of the three natural modes. By setting the out-of-plane amplitude to zero, critical points in the corresponding plane can be determined by analyzing the associated 2-DOF subsystem. The results, summarized in Table~\ref{tab2}, indicate that these additional critical points, whenever they exist, are always saddle points and hence correspond to dynamically unstable mixed-mode solutions that cannot be realized in the physical system.
\begin{table}[!htb]
\caption{Criteria for the existence of planar mixed-mode solutions in the 3-DOF system. The notation $\bar{A}_i$ denotes a non-zero steady-state amplitude for the respective mode.}
\label{tab2}
\centering
\renewcommand{\arraystretch}{1.25}
\begin{tabular}{@{}llll@{}}
\toprule
\textbf{Solution} &
$\left(\bar{A}_{1},\bar{A}_{2},0\right)$ &
$\left(\bar{A}_{1},0,\bar{A}_{3}\right)$ &
$\left(0,\bar{A}_{2},\bar{A}_{3}\right)$ \\
\midrule
\textbf{Criterion} &
$\strut 0.5 < q_{1} < 2$ &
$\strut 0.5 < q_{2} < 2$ &
$\strut 0.5 < \displaystyle \frac{q_{2}}{q_{1}} < 2$ \\
\bottomrule
\end{tabular}
\end{table}

\begin{figure}[!htb]
	\centering
	\includegraphics[width=\linewidth]{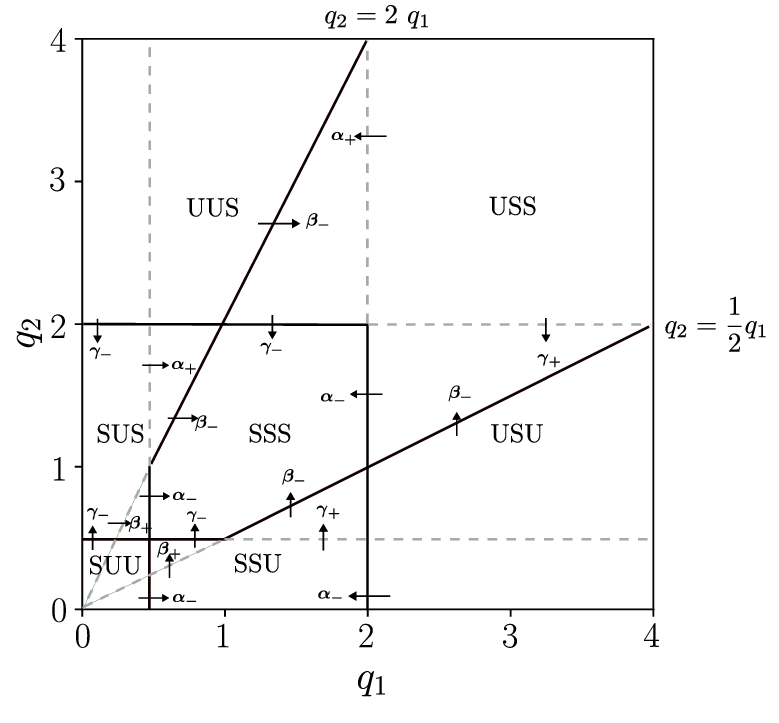}
    \caption{Slow-flow stability characteristics of the mixed-mode solutions in the $q_1$--$q_2$ parameter plane. Symbols $\alpha_{\pm}$, $\beta_{\pm}$, and $\gamma_{\pm}$ denote the planar saddle points at their inception, where the subscripts $-$ and $+$ identify attracting and repelling saddles, respectively. The birth of these points determines the global stability of the axial limit cycles, while the mixed-mode solutions themselves are always unstable in the 3DOF system}
	\label{fig:5}
\end{figure}

These saddle points emanate from the critical points on the axes through a subcritical pitchfork bifurcation, thereby altering the stability of the critical points on the axes within the corresponding coordinate planes. However, the out-of-plane eigenvalue associated with the axial critical points ultimately determines their stability in the full three-dimensional system. Consequently, these saddle points can exist even when only one of the modes is globally stable in the three-dimensional system, and their evolution can also be affected by the dynamics in the orthogonal planes. To completely characterize the evolution of the saddle critical points, we investigate the sign of their out-of-plane eigenvalue. For illustration, we consider a solution of the form $(\bar{A}_1,\bar{A}_2,0)$, and the expression for the out-of-plane eigenvalue of this critical point can be obtained as 
\begin{equation}
    \lambda_{3} = -\dfrac{\beta \omega_{3}^{2}}{2} + \dfrac{\beta \omega_{3}^{2}}{2\pi}  
    \begin{cases}
        \dfrac{\Phi_{33}\omega_{3}A_{3}^{*}}{\Phi_{31}\omega_{1}\bar{A}_{1}} \mathcal{K} \left( \delta_{1}^2 \right) &\text{if } \delta_{1} < 1 \\[2ex]
        \dfrac{\Phi_{33}\omega_{3}A_{3}^{*}}{\Phi_{32}\omega_{2}\bar{A}_{2}} \mathcal{K} \left(\delta_{1}^{-2} \right) &\text{if } \delta_{1} > 1
    \end{cases},
    \label{Eq:9}
\end{equation}
where $ \delta_{1} = (\Phi_{32}\omega_{2}\bar{A}_{2})/(\Phi_{31}\omega_{1}\bar{A}_{1})$, and $A_3^*$ is the critical point that lies on the out-of-plane axis. We categorize these mixed-mode solutions based on the sign of this eigenvalue: an attracting saddle, denoted as $\alpha_{-}$, possesses a negative out-of-plane eigenvalue ($\lambda_3 < 0$), whereas a repelling saddle, denoted as $\alpha_{+}$, possesses a positive one ($\lambda_3 > 0$). In the case of $\alpha_{-}$, the stable manifold is a 2-D surface while the unstable manifold is a line/curve \cite{thompson2002nonlinear}. Conversely, for $\alpha_{+}$, the out-of-plane direction is unstable, resulting in a 1-D stable manifold and a 2-D unstable manifold. While the amplitude values $\bar{A}_{1}$ and $\bar{A}_{2}$ generally require numerical computation, the out-of-plane stability simplifies significantly in the limiting conditions, where the sign of the eigenvalue is given by
\begin{equation}
	\text{sign}(\lambda_3) = \begin{cases}
		\text{sign} \left(-2 + q_{2} \right) & \text{for } \delta_{1} \rightarrow 0 \\
		\text{sign} \left(-2 + \dfrac{q_{2}}{q_{1}} \right) & \text{for } \delta_{1} \rightarrow \infty \\
	\end{cases}.
    \label{eq:10}
\end{equation}
Table~\ref{tab2} and Eq.~(\ref{eq:10}) reveal that the existence of this particular saddle point is governed by one system parameter, $q_{1}$, while its out-of-plane stability is controlled by the second parameter, $q_{2}$. For instance, when $q_{1} = 0.5$, the saddle point emanates from the axial critical point $(0, A_{2}^*, 0)$. At this axial limit, where $\delta_1 \to \infty$, the sign of the out-of-plane eigenvalue is governed by the second case in Eq.~(\ref{eq:10}). Consequently, if $q_2 < 1$ at this boundary, an attracting saddle $\alpha_{-}$ is created and the critical point becomes globally stable; i.e., the status of the limit cycle in the second mode changes from unstable to stable in the 3DOF system. On the contrary, if $q_{2} > 1$, a repelling saddle $\alpha_{+}$ is born, which does not alter the global stability of the second mode. Similarly, when the saddle merges with the other axial critical point $(A_1^*,0,0)$ at $q_1=2$ (where $\delta_1 \to 0$), the global stability of the first mode is altered only if $q_2<2$ (first case in Eq.~\ref{eq:10}). This analysis extends to the other two coordinate planes, where $\beta_{-}$ and $\beta_{+}$ denote saddle points in the $A_{1}$--$A_{3}$ plane, while $\gamma_{-}$ and $\gamma_{+}$ denote those in the $A_{2}$--$A_{3}$ plane, respectively. The complete characterization of these saddle points is shown in Fig.~\ref{fig:5}.

There are scenarios where the saddle may be born with a zero eigenvalue in the out-of-plane direction. Furthermore, the eigenvalue in Eq.~(\ref{Eq:9}) only has a lower bound due to the dependence on the elliptic function $\mathcal{K}$. As a result, saddle points that are initially attracting can also transition to repelling saddles as the system parameters vary, and vice versa. When a critical point is born with a zero eigenvalue (at $q_{1} = 0.5$, $q_{2} = 1$), it bifurcates into two saddle points located in the $A_{1}$--$A_{2}$ and $A_{1}$--$A_{3}$ planes, respectively. However, when the nature of the saddle changes with parameter variation, there are no additional saddle points and it becomes necessary to consider critical points that lie within the full three-dimensional phase space to look for additional bifurcations.

\subsubsection{\textbf{Critical points in the phase space}}
\label{sec:3dof_sign_num_sim}
In the absence of closed-form slow-flow equations in the full phase space, developing analytical techniques to examine the existence and stability of critical points of the form $(\hat{A}_{1}, \hat{A}_{2}, \hat{A}_{3})$ is not feasible. Such critical points represent quasi-periodic oscillations involving three natural frequencies of the three modal coordinates. Nevertheless, we can show that any critical point in the amplitude phase space must lie within the following bounds (derivation provided in the Appendix \ref{App_A}):
\begin{align}
	0 \le & \, \hat{A}_{1}\, \le \dfrac{2\varepsilon\Phi_{31}}{\pi\beta\omega_{1}^3}\left( 1+\sqrt{1+q_{1}+q_{2}}\right) = L_{1}, \nonumber \\
	0 \le & \,\hat{A}_{2}\, \le \dfrac{2\varepsilon\Phi_{31}}{\pi\beta\omega_{1} \omega_2^2} \left(\sqrt{q_{1}}+\sqrt{1+q_{1}+q_{2}}\right) < 2L_{1}, \nonumber \\
	0 \le & \, \hat{A}_{3}\, \le \dfrac{2\varepsilon\Phi_{31}}{\pi\beta\omega_{1} \omega_3^2} \left( \sqrt{q_{2}}+\sqrt{1+q_{1}+q_{2}}\right) <  2L_{1}. 
	\label{eq:11}		
\end{align}
Since the domain of existence of critical points is bounded, numerical simulations can be performed in the physical coordinates to investigate the global stability of the system. Because three single-mode limit cycles always exist, and one of them is always stable, the entire phase space is partitioned into basins of attraction for each mode. Consequently, if stable mixed-mode solutions also existed in the slow-flow phase space, a subset of initial conditions would necessarily converge to them. The following algorithm is employed to investigate the presence of stable mixed-mode oscillations using numerical simulations:
\begin{itemize} 
    \item \textbf{Initial Condition Distribution:} For each point in the $q_1$--$q_2$ parameter space, a set of initial conditions contained within a ball of radius $2L_1$ is considered. Half of the initial conditions are sampled near the origin (the unstable trivial solution), while the remaining half are distributed near the boundary specified in Eq.~(\ref{eq:11}), ensuring comprehensive coverage of the phase space. 
    \item \textbf{Steady-State Compilation:} For each initial condition, the physical system is simulated until transients decay. The resulting steady-state amplitudes are recorded, and the results are compiled as a percentage of initial conditions converging to each specific limit cycle mode. 
\end{itemize}

\begin{figure*}[!htb]
	\centering
	\includegraphics[width=.75\linewidth]{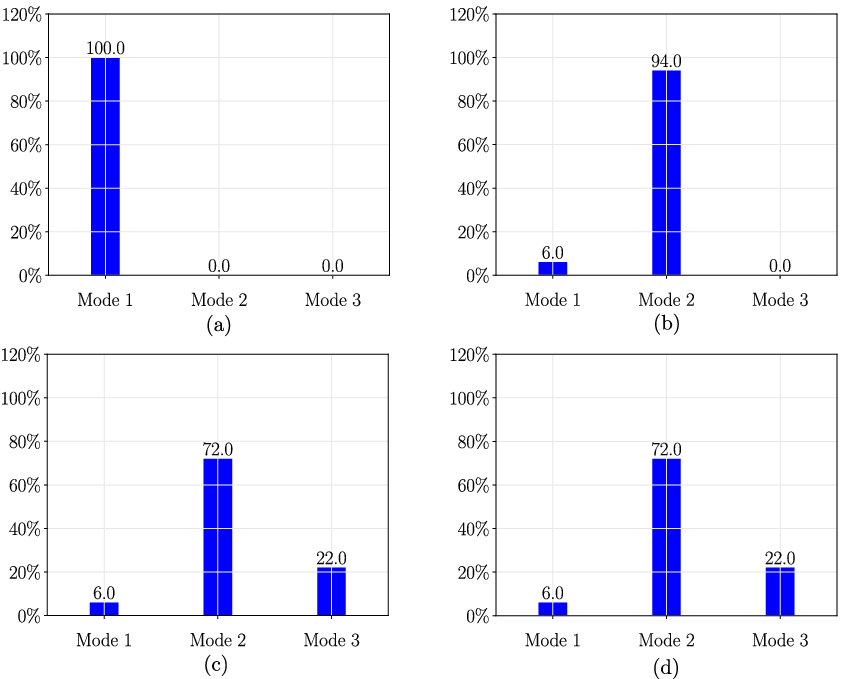}
	\caption{Numerical analysis of the steady-state dynamics of the 3-DOF system for varying parameter configurations. Each panel illustrates the percentage of initial conditions attracted to limit cycle oscillations across all modal coordinates; (a) $~q_1=0.1,\; q_2=0.1 $ (SUU), (b) $ q_1=1.067,\; q_2=0.422 $ (SSU), (c) $ q_1=1.14,\; q_2=1.19 $ (SSS), (d) $ q_1=1.14,\; q_2=1.191  $ (SSS)}
	\label{fig:6}
\end{figure*}

The results of numerical simulations for three representative choices of $q_{1}$ and $q_{2}$, corresponding to the SUU, SSU, and SSS stability categories described earlier, are shown in Fig.~\ref{fig:6}(a)-(c). For the SUU case, all initial conditions converge to the limit cycle with amplitude and frequency corresponding to the first natural mode of oscillation. For the other two cases, all initial conditions converge to a limit cycle in one of the fundamental modes of oscillation, and no additional steady-state solutions are observed. Similarly, simulations for four other parameter combinations, corresponding to the remaining stability categories (presented in the Appendix \ref{App_B} for completeness), confirm the same observation: no additional steady-state solutions are observed.

The parameters for the SSS scenario in Fig.~\ref{fig:6}(c) are chosen such that the out-of-plane eigenvalue of one of planar saddle points is close to zero. Consequently, a small variation in one parameter induces a qualitative change in the nature of the saddle point. Figure~\ref{fig:6}(d) illustrates that this change in saddle stability does not affect the global dynamics of the system; all initial conditions still converge to one of the three fundamental modes of oscillation. These numerical results demonstrate that such qualitative changes in the nature of the saddle critical points do not influence the global stability of the system. In other words, the overall dynamics remain governed by the axial critical points and their associated bifurcations. 

To verify that the numerical observations in Fig.~\ref{fig:6}(a)--(c) are topologically consistent across a wider set of parameters, simulations were performed over an extended $q_{1}$--$q_{2}$ domain ranging from $0.1$ to $3$. The system parameters were chosen to avoid repeated frequencies and internal resonance conditions, thereby precluding complex modal interactions that could obscure the primary stability transitions. Highly skewed mass and stiffness distributions, which are predominantly observed when $q_1$ and $q_2$ exceed 3, were excluded from this analysis as they do not represent physically meaningful configurations. In all cases, the simulations converged to a steady-state limit cycle, with the frequencies matching those of one of the natural modes of the system. Additionally, within the physically relevant parameter range, mixed-mode saddle points do not bifurcate to produce global attractors; asymptotically, trajectories converge to single-mode limit cycles aligned with the axial critical points. Further, there are also no other bifurcations (local \& global) that are observed in the system.

\begin{figure*}[!htb]
	\centering
	\includegraphics[width=1\linewidth]{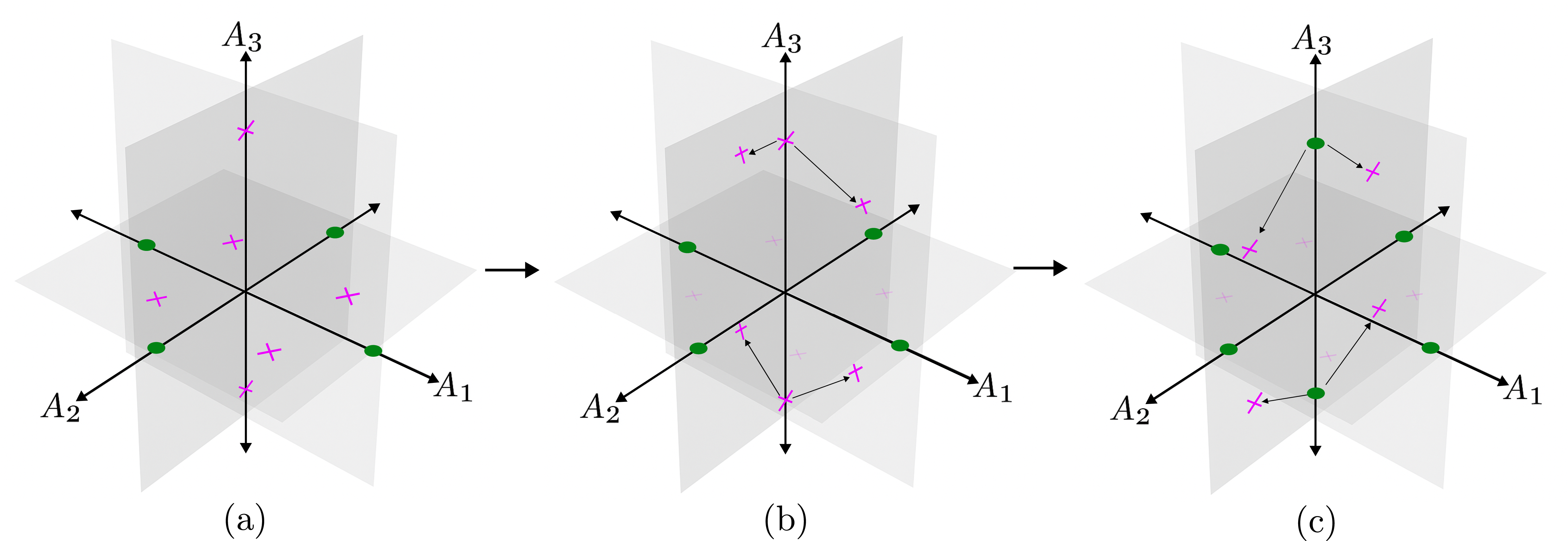}
    \caption{Bifurcation sequence for the 3-DOF system illustrating the transition of an axial critical point from SSU to SSS for $q_1 \in \{1.0, 2.0\}$, with $q_2$ acting as the bifurcation parameter}
	\label{fig:7}
\end{figure*}

In summary, even in the 3-DOF system, the steady-state response always corresponds to a single-mode limit cycle. Any qualitative changes in the phase space are captured by a series of subcritical pitchfork bifurcations. The in-plane transition, such as the shift from SUU to SSU, occurs when an axial critical point changes stability and sheds two attracting saddles within the specific coordinate plane. This mirrors the 2-DOF behavior shown in Fig.~\ref{fig:3}. The more complex progression toward a tristable SSS configuration involves a series of bifurcations of the same critical point across different planes. This process is illustrated in Fig.~\ref{fig:7} for the transition from SSU to SSS where $q_1 \in [1,2]$. At $q_2=0.5$, the axial critical point $\pm (0,0,A_3^*)$ splits into saddle points within the $A_1$--$A_3$ plane. Upon crossing the (solid) boundary at $q_2=0.5q_1$, it splits further into the $A_2$--$A_3$ plane while altering its stability. Finally, as $q_2$ exceeds 2, the $A_1$--$A_3$ saddles merge with the $A_1$ axial nodes, causing the first mode to lose stability. This completes the generalized ``SAF'' (Stability-Axis-Flipping) sequence, moving the system from a configuration where mode 1 was stable (SSU) to one where it is unstable (USS). As in the 2-DOF case, this mechanism is defined by the migration of saddle points that mediate the exchange of stability between fundamental modes in a higher-dimensional phase space.

While extensions to higher degrees of freedom are in principle possible, the implicit forcing used in this study excites all natural modes irrespective of their modal damping, thereby generating limit cycles in each mode. Since this behavior may not reflect more realistic physical scenarios, we next consider a modified implicit forcing that incorporates an explicit dependence on the modal damping in the system.

\section{Modified Mathematical Model}\label{section_3}

The forcing applied to the terminal mass in Fig.~\ref{fig:1} is modified to incorporate negative damping and a velocity-weighted discontinuous component defined as
\begin{equation*}
    F(x,\dot{x})
    \;=\;\varepsilon\,\dot{x}\,H(x_{0}+x)
    \;=\;
    \tilde{\varepsilon}\,\dot{x}
    \;+\;
    \tilde{\varepsilon}\,\dot{x}\,\text{sign}(x_{0}+x)   
%    \label{eq17}
\end{equation*}
where $\tilde{\varepsilon}=\varepsilon/2$ and $H(\cdot)$ is the Heaviside function. The parameter $x_{0}$=$\mathcal{O}(\varepsilon)$ introduces a bias that breaks the symmetry of the switching surface. This forcing makes explicit that small, velocity-dependent perturbations arise even near zero displacement and are inherently destabilizing when activated (for $x>-x_{0}$). The proposed formulation leads to a piecewise-linear dynamical system, variations of which have been employed to investigate bifurcation phenomena in planar systems~\cite{leine2013dynamics}. 

The analysis of the corresponding SDOF system (Appendix \ref{App_C}) reveals that the qualitative response is governed by the interplay between the forcing magnitude ($\varepsilon$)  and the intrinsic system parameters. Specifically, the stability of the equilibrium depends on whether $\varepsilon$ exceeds a critical threshold defined by the natural dissipation; depending on this balance, the system may behave as a damped harmonic oscillator, sustain a stable limit cycle, or exhibit unbounded growth. Unbounded growth is atypical; throughout this work, we restrict attention to parameter values for which trajectories remain bounded. As a modeling note, the unstable region can also be eliminated by adopting a two-parameter weak forcing (e.g., $\varepsilon_1$, $\varepsilon_2$) with appropriately constrained values. In MDOF configurations, this effective negative damping competes with modal damping, allowing the forcing to selectively destabilize specific modes based on their modal projections and inherent damping levels.

\subsection{Four-degree-of-freedom (4-DOF) system}
We consider a system with four degrees of freedom subjected to the Heaviside forcing function. Using the method of averaging, the amplitude of each mode of oscillation can be obtained as
\begin{align}
\dot{A}_{i} &= -\frac{\beta \omega_{i}^{2} A_{i}}{2}
+ Q_{i},~i=1,2,3,4, \label{eq:12}
\end{align}
\text{where}
\begin{align*}
Q_{i} &= \frac{\epsilon \Phi_{4i}^{2}}{(2\pi)^4\omega_{i}} \int_{0}^{2\pi}\!\cdots\!\int_{0}^{2\pi}
\tilde{Q}_{i} \, d\theta_{1}\cdots d\theta_{4}, \nonumber \\
\tilde{Q}_{i} &= \dot{x}_4\,\mathrm{H}\!\left(x_{0} + x_4\right)\sin\theta_{i} \nonumber \\
x_4 & = \sum_{j=1}^{4} \Phi_{4j}A_{j}(t)\cos\theta_{j}(t). \nonumber 
\end{align*}
The integrals in Eq.~(\ref{eq:12}) do not admit closed-form expressions even for a two-degree-of-freedom system due to the offset $x_{0}$. Consequently, explicit slow-flow equations cannot be derived for any system with more than one degree of freedom. Nevertheless, by replicating the analysis performed for the 2-DOF and 3-DOF systems in section~\ref{Sign}, the dynamics of the 4-DOF configuration can be evaluated as a function of the system parameters and forcing magnitude (given by $\tilde{q}_{i}$). In particular, a limit cycle in the $k^{\text{th}}$ mode ($k\le 4$) exists depending on the parameter combinations listed in Table~\ref{tab3}.
\begin{table}[!ht]
\caption{Stability criteria for the single-mode limit cycles in the 4-DOF system. The parameter $\tilde{q}_i = (\Phi_{4i}^2\varepsilon)/(\beta\omega_{i}^2)$ represents the normalized ratio of energy input to modal damping for the $i^{\text{th}}$ mode}
\label{tab3}
\centering
\renewcommand{\arraystretch}{1.25}
\begin{tabular}{@{}ll@{}}
\toprule
\textbf{Mode} & \textbf{Criteria for stability} \\
\midrule
$1^{\text{st}}$ mode $(A_1^*,0,0,0)$ &
$\strut 1 < \tilde{q}_{1} < 2$ \\

$2^{\text{nd}}$ mode $(0,A_2^*,0,0)$ &
$\strut 1 < \tilde{q}_{2} < 2$  \\

$3^{\text{rd}}$ mode $(0,0,A_3^*,0)$ &
$\strut 1 < \tilde{q}_{3} < 2$  \\

$4^{\text{th}}$ mode $(0,0,0,A_4^*)$ &
$\strut 1 < \tilde{q}_{4} < 2$  \\
\bottomrule
\end{tabular}
\end{table}

\begin{figure*}[!htb]
	\centering
	\includegraphics[width=.75\linewidth]{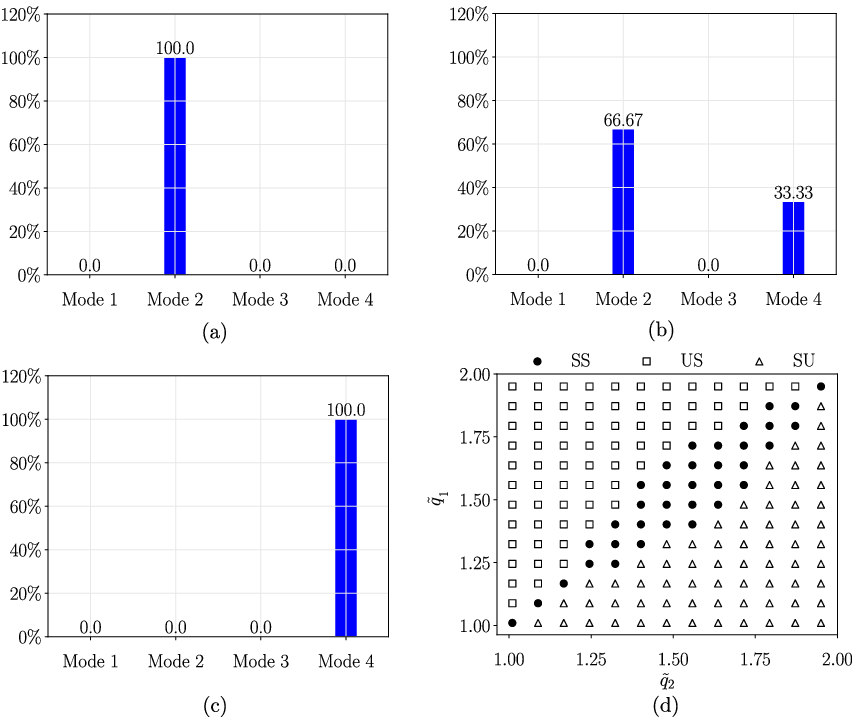}
	\caption{Numerical analysis of the steady-state dynamics of the 4-DOF system under Heaviside forcing for specific values of $(\tilde{q}_2, \tilde{q}_4)$, with $\tilde{q}_{1}, \tilde{q}_{3} < 1$. (a) $\tilde{q}_2 = 1.66,~\tilde{q}_4=1.23$, (b) $\tilde{q}_2 = 1.55,~\tilde{q}_4=1.55$, (c) $\tilde{q}_2 = 1.23,~\tilde{q}_4=1.77$, (d) Stability map of the limit cycle oscillations in the $q_2$--$q_4$ parameter space}
	\label{fig:8}
\end{figure*}

For illustration, consider a system for which $\tilde{q}_1,\tilde{q}_3<1$ and $1<\tilde{q}_2,\tilde{q}_4<2$, implying that limit cycles can exist only in the second and fourth natural modes. In this case, the analysis reduces to a 2-DOF system in modes 2 and 4, and the critical eigenvalues of the root $(0,A_2^*,0,0)$ can be obtained as
\begin{align*}	
    \lambda_{2} &= - \dfrac{\beta \omega_{2}^2}{2} \left( 1  - \dfrac{\tilde{q}_{2}}{\pi} \left( \pi - \arccos{\left( \hat{\delta_2} \right)} - f_{7}\right) \right) \nonumber   \\ 
    \lambda_{4} &= -\frac{\beta\omega_{4}^2}{2} \left(1 - \dfrac{\tilde{q}_{4}}{\pi} \left(\pi - \arccos (\hat{\delta}_2)\right) \right),	
\end{align*}
where
\begin{align*}
\hat{\delta}_2 = x_0/(\Phi_{42}A_2^*) \text{ and } f_{7} = \hat{\delta}_2 \sqrt{1-\hat{\delta}_2^2 }.
\end{align*}
$\lambda_{2}$ represents the equivalent SDOF eigenvalue for the second mode and therefore remains negative for $\tilde{q}_{2}<2$. For the other eigenvalue ($\lambda_4$) to be negative, we require
\begin{equation}
    \dfrac{\tilde{q}_{4}}{\pi}\,\arccos\left( -\hat{\delta}_2\right) < 1.
    \label{eq:13}
\end{equation}
Thus, even when the limit cycle oscillation is stable in the corresponding SDOF setting, mode coupling can induce a loss of stability in the coupled system. Similarly, the axial critical point $(0,0,0,A_4^*)$ is stable when
\begin{equation}
    \dfrac{\tilde{q}_{2}}{\pi}\,\arccos\left( -\hat{\delta}_4\right) < 1.
    \label{eq:14}
\end{equation}
where $\hat{\delta}_4 = x_{0}/(\Phi_{44}A_{4}^{*})$. Thus, both axial limit cycles are stable when Eqs.~(\ref{eq:13}) and ~(\ref{eq:14}) are satisfied. In this case, at least one additional saddle point must exist in the slow-flow plane to partition the phase space into the basins of attraction of the two stable nodes. While the transcendental nature of the equations precludes a closed-form proof of uniqueness for this saddle, numerical simulations for all tested combinations of $(\tilde{q}_{2},\tilde{q}_{4})$ confirm that trajectories converge exclusively to axial limit cycles. As illustrated in Fig.~\ref{fig:8}(a)–(c), no evidence of stable mixed-mode attractors or secondary bifurcations was observed within the investigated parameter range.

In summary, in this MDOF system, limit cycles are possible in the second and fourth modes, with stability governed by the specific choices of $\tilde{q}_{2}$ and $\tilde{q}_{4}$, as illustrated in Fig.~\ref{fig:8}(d). Accordingly, the steady-state dynamics may consist of a limit cycle in either mode, or exhibit bistability in which the steady-state solution depends on the initial conditions.

The preservation of the stability-axis-flipping (SAF) bifurcation mechanism across both forcing models is consistent with the inherent mathematical relationship between the Sign and Heaviside functions. Despite the introduction of switching asymmetry ($x_0$) and state-dependent weighting in the modified model, the global dynamics remain qualitatively anchored to the natural modes. This indicates that the parameter-driven exchange of stability mediated by mixed-mode saddles is a robust feature of this class of non-smooth excitation, surviving even when the forcing symmetry is broken.

In conclusion, the steady-state behavior of the MDOF system under discontinuous, state-dependent forcing with an offset and a destabilizing term remains qualitatively similar to the earlier cases. Mode coupling can induce a loss of stability in the fundamental mode, leading to a stable limit cycle in a higher mode, or result in bistable responses where the final limit cycle depends on initial conditions or small perturbations.

\section{Conclusion}\label{Conclusion}

This study investigates the existence and stability of limit cycles in linearly damped MDOF systems subjected to discontinuous, state-dependent feedback mechanisms. By utilizing the method of averaging and slow-flow phase-plane analysis, this work maps the stability boundaries of periodic responses within the amplitude space. Furthermore, the analysis eliminates the possibility of quasi-periodic responses through rigorous qualitative arguments and numerical simulations.

For the symmetric relay feedback analyzed in Section \ref{Sign},  analysis for a 2-DOF system demonstrates that a stable limit cycle always exists. The analysis shows that mode coupling can result in a bistable regime where both modal limit cycles are stable and the final steady-state response depends on the initial conditions. As parameters vary, this coupling eventually destabilizes the limit cycle in the fundamental mode and forces the system to settle onto a stable limit cycle in the higher mode. This transition is defined by a stability-axis-flipping (SAF) bifurcation, which serves as the governing mechanism even as the degrees of freedom are increased. In higher-order systems, the limit cycles continue to exchange stability through a higher-dimensional analogue of the SAF bifurcation.

The investigation of the additional feedback mechanism in Section \ref{section_3} yields closed-form criteria for the SDOF case and establishes the generality of the framework using a 4-DOF system. These findings confirm that the system response remains fundamentally consistent, as mode coupling destabilizes the fundamental mode and facilitates the emergence of stable limit cycles in higher modes. Because the SAF bifurcation also governs this transition, the results establish its role as a universal feature within a broader class of piecewise-linear dynamical systems.

In summary, this work provides a systematic analytical framework for the steady-state dynamics of MDOF systems subjected to state-dependent discontinuous feedback. The findings integrate into stability maps that define the criteria for the existence and stability of limit cycles across different modes. By circumventing the need for extensive brute-force numerical simulations, this framework offers an efficient approach for guiding optimization studies to mitigate or generate limit cycles at targeted frequencies. Future expansions of this framework will incorporate sliding mode dynamics to capture the essential features of stick-slip phenomena. Establishing qualitative parallels between these analytical models and empirical field data will yield a robust mathematical framework to predict stick-slip instabilities across higher modes in realistic systems.

\section*{Acknowledgment} %% ASME requests this exact spelling, singular.

\noindent The authors are grateful to the Industrial Research and Consultancy Centre (IRCC) of the Indian Institute of Technology Bombay for funding this research. The authors also acknowledge the institutional support received from the Indian Institute of Technology Bombay.

\section*{Author contributions} 
Both authors contributed to the conceptualization and methodology of the research. Arunav Choudhury carried out the formal mathematical analysis and numerical simulations, generated the figures, and prepared the original manuscript draft. R. Ganesh contributed to the mathematical analysis, supervised the research, provided resources, and reviewed and edited the manuscript. Both authors discussed the results and approved the final manuscript.

\section*{Conflict of Interest}

The authors declare that they have no known competing financial interests or personal relationships that could have appeared to influence the work reported in this paper.

%%%%%%------NONEMCLATURE (checking)-------
\appendix

%% ---------------------------------------------------%%%
%%------FOR continuation of equation, figure and table numbers w.r.t. main body and removng A.,B.,C.-----%%
%%-- THIS IS STATAED IN ASME GUIDELINES-----%%
\renewcommand{\theHequation}{appendix.\arabic{equation}} % to prevent hyperlinks from jumping to the wrong equation number
\makeatletter
% 1. Restore the equation number
\setcounter{equation}{14} % Or whatever your last number was
% 2. Remove the "A." prefix for the label
\renewcommand{\theequation}{\arabic{equation}} 
% 4. Overwrite the template's "reset-to-zero" hook
\titleformat{\section}[block]{\mathversion{bold}\bfseries\large\raggedright}{\appendixname\ \thesection:}{0.5em}{}
\makeatother
%---------------------------------------------------

\section{Domain for critical points in the 3-DOF system}
\label{App_A}

For the 3-DOF system considered in section~\ref{sec:3dof_sign_num_sim}, the averaged equations can be written as 
\begin{align*}
	\dot{A}_{i} &= -\dfrac{\beta \omega_{i}^2 A_{i}}{2} 
	+ \dfrac{\epsilon \Phi_{3 i}}{8\pi^{3}\omega_{i}} 
	T_{i},\quad i = 1,2,3. \nonumber \\
    T_{i} &= \int_{0}^{2\pi}\int_{0}^{2\pi} \int_{0}^{2\pi} 
	 \tilde{T}_{i} \, d\theta_{1}d\theta_{2}d\theta_{3}, \\
     \tilde{T}_{i} & = \text{sign}\left(\sum_{j=1}^{3}\Phi_{3j}\omega_{j}A_{j}\sin \theta_{j} \right) 
	 \sin \theta_{i}.
\end{align*}
Using the fact that $\dot{A}_{i}=0$ for a critical point, the individual equations can be rearranged to obtain
\begin{align*}
	& \sum_{i=1}^{3} \omega_{i}^{4}A_{i}^{2}  = \dfrac{\epsilon}{4\pi^{3}\beta} S,\\
    & S  = \int_{0}^{2\pi}\int_{0}^{2\pi} \int_{0}^{2\pi} \left \lvert  \sum_{i=1}^{3}\Phi_{3i}\omega_{i}A_{i}\sin \theta_{i} \right \rvert \, d\theta_{1}d\theta_{2}d\theta_{3} 
\end{align*}
Using the triangle inequality, the integral term can be simplified as 
\begin{align}
S & \le \int_{0}^{2\pi}\!\!\int_{0}^{2\pi}\!\!\int_{0}^{2\pi}
\sum_{i=1}^{3} \Phi_{3i}\omega_{i}A_{i}\lvert \sin \theta_{i} \rvert
\, d\theta_{1} d\theta_{2} d\theta_{3} \nonumber \\
  &\le \sum_{i=1}^{3} 16\pi^{2}\Phi_{3i}\omega_{i}A_{i} \nonumber \\
\implies
&\sum_{i=1}^{3}\omega_{i}^{4}A_{i}^{2}
  \le \dfrac{4\epsilon}{\pi\beta}
     \sum_{i=1}^{3} \Phi_{3i}\omega_{i}A_{i} \nonumber 
  \le \sum_{i=1}^{3} \omega_{i}^{4} A_{i}^{*}A_{i} \nonumber \\
 \implies &
\sum_{i=1}^{3} \omega_{i}^{4}A_{i}\left(A_{i}-A_{i}^{*}\right)
  \le 0 \nonumber
\end{align}
\noindent
where $A_i^*$ denote the axial critical points given in section \ref{cr_pts_axes}. Thus, non-trivial solutions of the form $(\hat{A}_{1},\hat{A}_{2},\hat{A}_{3})$ can exist if
\begin{equation*}
	\hat{A}_{1} < A_{1}^{*},~ \hat{A}_{2} < A_{2}^{*}, \text { and } \hat{A}_{3} < A_{3}^{*}. 
\end{equation*}
In other words, a mixed-mode solution can exist if all modal amplitudes are bounded by their respective axial critical points. However, to find the absolute global bounds, we consider the case where one amplitude may exceed its axial value at the expense of others. If $\hat{A}_{1} > A_{1}^{*}$ and $ \hat{A}_{2} < A_{2}^{*},\, \hat{A}_{3} < A_{3}^{*} $, then we have
\begin{align}
\hat{A}_{1}\left( \hat{A}_{1} - A_{1}^{*} \right) \leq  \left( \dfrac{\omega_{2}}{\omega_{1}} \right)^{4} \hat{A}_{2}\left( \hat{A}_{2}^* - A_{2} \right) + \nonumber \\ \left( \dfrac{\omega_{3}}{\omega_{1}} \right)^{4}  \hat{A}_{3}\left( \hat{A}_{3}^* - A_{3} \right). \label{eq:15}
\end{align}
The maximum value of R.H.S. in Eq.~(\ref{eq:15}) occurs at $\hat{A}_{2}=0.5A_{2}^{*}$ and $\hat{A}_{3}=0.5 A_{3}^{*}$. Therefore, we have
\begin{equation*}
	\dfrac{\hat{A}_{1}}{A_{1}^{*}} \left(\dfrac{\hat{A}_{1}}{A_{1}^{*}} - 1\right) \leq \dfrac{\omega_{2}^{4}\left( A_{2}^{*}\right)^{2}+\omega_{3}^{4}\left( A_{3}^{*}\right)^{2}}{4\left(A_{1}^{*} \right)^{2}\omega_{1}^{4}}.
\end{equation*}
Substituting the values of $A_{2}^{*}$, $A_{3}^{*}$ and simplifying, we obtain
\begin{equation*}
	\dfrac{\hat{A}_{1}}{A_{1}^{*}} \left(\dfrac{\hat{A}_{1}}{A_{1}^{*}} - 1\right) \leq \dfrac{\left(q_{1}+q_{2} \right)}{4}.
\end{equation*}
Solving for the positive root of this quadratic equation yields the upper bound $L_1$ defined in Eq.~(\ref{eq:11}). This analysis can be repeated for $\hat{A}_{2}$ and $\hat{A}_{3}$ to establish the respective bounds $L_{2}$ and $L_{3}$, given as 
\begin{align*}
    L_{1} & =\dfrac{2\varepsilon\Phi_{31}}{\pi\beta\omega_{1}^3}\left( 1+\sqrt{1+q_{1}+q_{2}}\right), \\
    L_{2} &=\dfrac{2\varepsilon\Phi_{31}}{\pi\beta\omega_{1}^3 }\left(\dfrac{\omega_{1}}{\omega_{2}} \right)^2 \left(\sqrt{q_{1}}+\sqrt{1+q_{1}+q_{2}}\right), \\
    L_{3} & = \dfrac{2\varepsilon\Phi_{31}}{\pi\beta\omega_{1}^3 }\left(\dfrac{\omega_{1}}{\omega_{3}} \right)^2 \left( \sqrt{q_{2}}+\sqrt{1+q_{1}+q_{2}}\right). 
\end{align*}
To evaluate the upper bound for simulations, consider the ratio 
\begin{align*}
    \frac{L_{2}}{L_{1}} &= \left( \frac{\omega_{1}}{\omega_{2}} \right)^{2} \frac{\sqrt{q_{1}} + \sqrt{1 + q_{1} + q_{2}}}{1 + \sqrt{1 + q_{1} + q_{2}}} \\
    &= \left( \frac{\omega_{1}}{\omega_{2}} \right)^{2} \left(1+\frac{\sqrt{q_{1}} - 1}{1 + \sqrt{1 + q_{1} + q_{2}}}\right).
\end{align*}
Since $q_{1}, q_{2} > 0$, the ratio can be simplified as 
\[\frac{L_{2}}{L_{1}} < 2 \left( \frac{\omega_{1}}{\omega_{2}} \right)^{2}. \]
Similarly, we can also obtain
\[\frac{L_{3}}{L_{1}} < 2 \left( \frac{\omega_{1}}{\omega_{3}} \right)^{2}. \]
Assuming, without loss of generality, that $\omega_1 \le \omega_2 \le \omega_3$, the upper bound for $L_2$ and $L_3$ are established as $2L_1$. 
\section{Numerical analysis of 3-DOF system subjected to sign forcing}
\label{App_B}
\begin{figure*}[!htb]
	\centering
	\includegraphics[width=.75\linewidth]{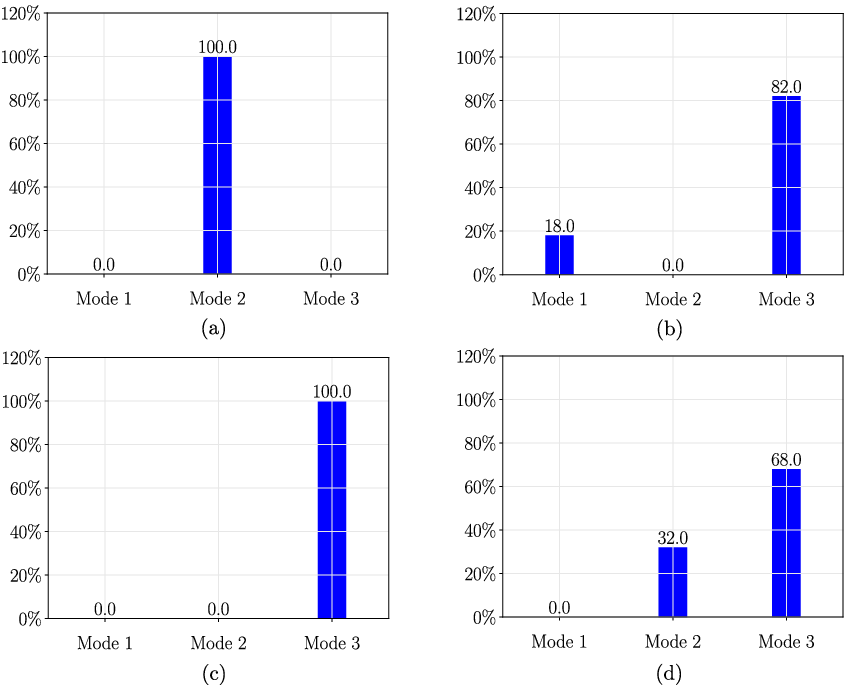}
	\caption{Steady-state dynamics for the remaining 3-DOF stability categories. In all cases, every sampled trajectory converges to a single-mode limit cycle, confirming that the axial critical points remain the only global attractors across these parameter regimes; (a) $q_{1}=2.678,~q_{2}=0.744$ (USU), (b) $q_{1}=0.1,~q_{2}=1.389$ (SUS), (c) $q_{1}=0.422,~q_{2}=3$ (UUS), (d) $q_{1}=1.711,~q_{2}=3$ (USS)}
	\label{fig:9}
\end{figure*}

The steady-state convergence for the remaining four stability configurations (UUS, USU, SUS, and USS) is shown in Fig.~\ref{fig:9}. These results confirm that the global dynamics remain consistent across the entire parameter space: all trajectories converge to fundamental axial modes, and no stable mixed-mode attractors are observed.

\section{Analysis of SDOF system subjected to Heaviside forcing}
\label{App_C}

\begin{figure*}[!htb]
	\centering
	\includegraphics[width=.75\linewidth]{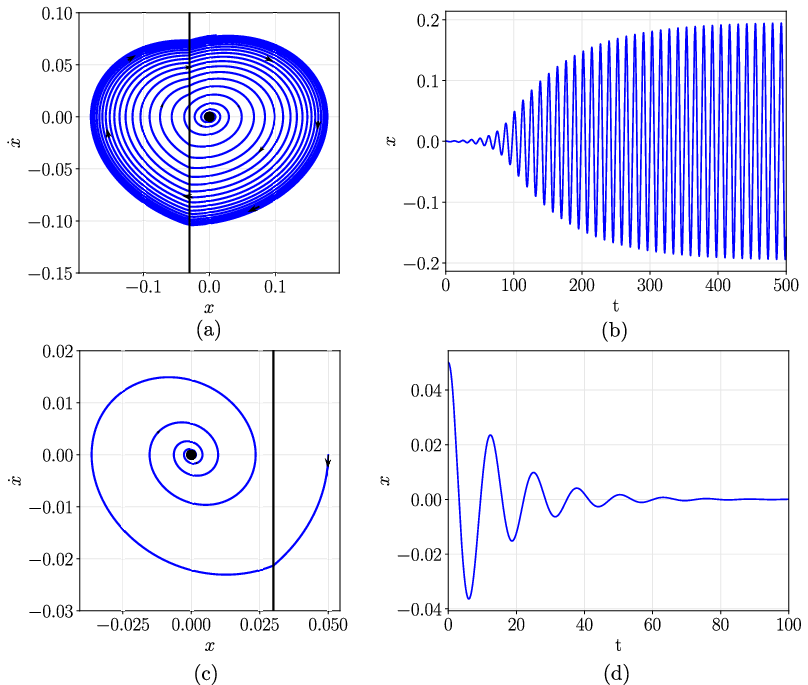}
		\caption{Numerical simulation of the SDOF system under Heaviside forcing ($\tilde{q} = 1.67$): (a)-(b) for $x_0 = 0.03$, the equilibrium at the origin is unstable, resulting in convergence to a stable limit cycle; (c)-(d) for $x_0 = -0.03$, the origin lies in the dissipative region, and all trajectories decay to the trivial steady state solution}
	\label{fig:10}
\end{figure*}

The equation of motion for the SDOF system can be written as:
\begin{equation}
	\ddot{x}+\beta \omega^2 \dot{x}+\omega^2 x = \epsilon\,  \dot{x}\, \text{H}(x_{0}+x). 
	\label{eq:16}
\end{equation}
 When $\beta<\epsilon$, there is addition of energy in the region where $x_{0}+x>0$, and removal of energy when $x_{0}+x<0$. The relative magnitudes of the energy addition/removal depends on the sign of the parameter $x_0$. This is illustrated in Fig.~\ref{fig:10}, where the results of the numerical simulation are demonstrated for $x_{0}>0$ and $x_0<0$. In Fig.~\ref{fig:10}(a), when $x_0>0$, the critical point $(x^{*},\dot{x}^{*})=(0,0)$ lies in the region where energy is supplied to the system and is unstable ($Re(\lambda_{(0,0)})>0$). When an initial condition lies in the vicinity of this critical point, the system evolves upon gaining energy until it crosses over onto the region where energy is dissipated. Depending upon the relative strengths of the damping in these two regions, the system may exhibit a periodic cycle, or unbounded growth (not shown here for brevity). However, when $x_{0}<0$, the critical point $(0,0)$ lies in the region where energy is dissipated (Fig.~\ref{fig:10}(c)) i.e., $Re(\lambda_{(0,0)})<0$. Hence, there is no growth in the amplitude and for all initial conditions lying in the region of dissipation, oscillations decay. Even when the initial conditions lie in the region where $x>x_{0}$, the oscillations decay. Therefore, when $x_{0}$ changes sign, the eigenvalues of the critical point change its stability, i.e., the system undergoes a Hopf bifurcation \cite{thompson2002nonlinear}. The eigenvalues jump from the left side of the complex plane to the right side when the parameter $x_{0}$ crosses $0$ to become positive.

\begin{figure}[!htb]
	\centering
	\includegraphics[width=.85\linewidth]{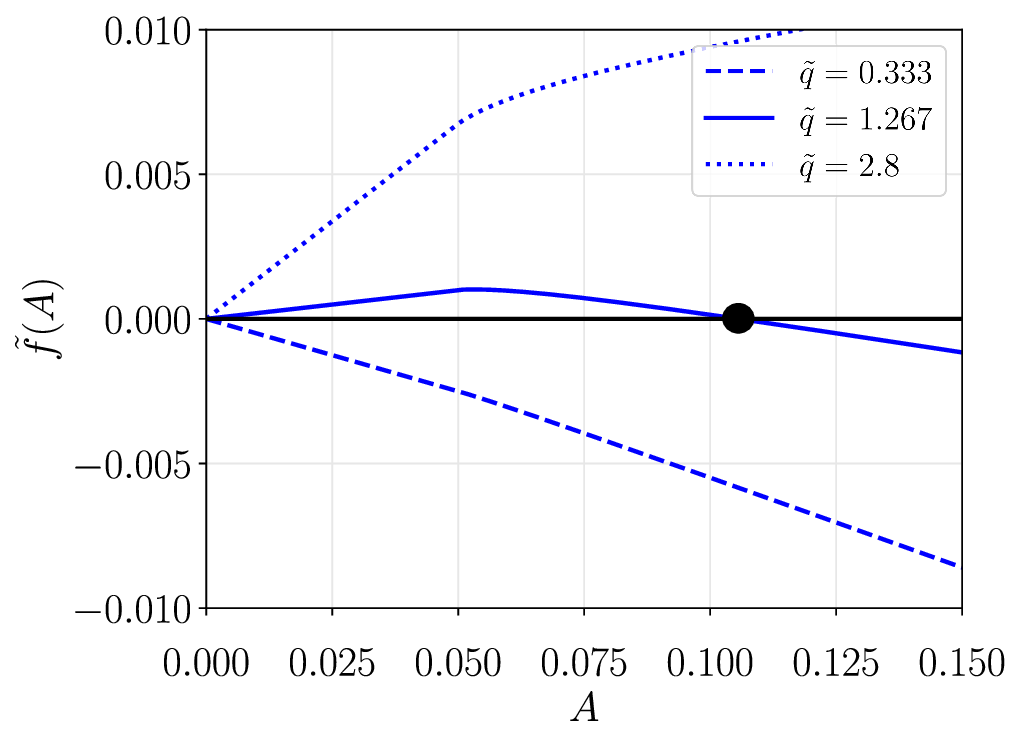}
    \caption{Graphical analysis of the slow-flow function $\tilde{f}(A)$ for varying $\tilde{q}$. The trivial solution always exists and is unstable for $\tilde{q}>1$. When $1<\tilde{q}<2$, an additional stable root also exists, which corresponds to a limit cycle in the physical system coordinates}
    \label{fig:11}
\end{figure}

Since the discontinuous forcing includes a spatial offset, the potential for sliding modes must be addressed. However, the interaction of the vector field with the switching surface $h(x, \dot{x}) = x + x_0$ is governed by the projections of the local flows onto the surface normal. The transversality condition for the absence of sliding is given as~\cite{leine2013dynamics}
\begin{equation}
    (\nabla h \cdot \mathbf{F}_1) (\nabla h \cdot \mathbf{F}_2) > 0,
    \label{eq:17}
\end{equation}
where $\mathbf{F}_1$ and $\mathbf{F}_2$ represent the vector fields on either side of the discontinuity, and $\nabla h = [1, 0]^T$ is the gradient of the switching surface in the $(x, \dot{x})$ phase plane. For this system, the condition in Eq.~\eqref{eq:17} is satisfied for all non-zero velocities, precluding the existence of a sliding mode. This allows for an eigenvalue-based stability analysis and the application of the method of averaging to determine the slow-flow dynamics. The slow-flow equation governing the evolution of the oscillation envelope (A) for Eq.~(\ref{eq:16}) with $x_0>0$ can be obtained as
\begin{equation}
	\dot{A} = \tilde{f}(A) = 
	\begin{cases} 
		\dfrac{\beta\omega^2 A}{2} \left(-1  + \tilde{q}\right) , 
		& \text{if } A <x_0, \\		
		\dfrac{\beta\omega^2 A}{2} \left[
		-1 + \dfrac{\tilde{q}}{\pi}\, 
			f_{8} \right], & \text{if } A>x_0
	\end{cases}
	\label{eq:18}
\end{equation}
where
\begin{align*}
    f_{8} &= \left( \pi - \arccos \left( \hat{\delta} \right) + \hat{\delta} \sqrt{1 - \left( \hat{\delta} \right)^2} \right), \\
    \hat{\delta} &= \dfrac{x_0}{A},\ \tilde{q} =\dfrac{\epsilon}{\beta \omega^2}.
\end{align*}
The parameter $\tilde{q}$ represents the normalized ratio of energy input to modal damping. The trivial solution is stable if $\tilde{q}<1$ and unstable if $\tilde{q}>1$. The non-trivial solutions satisfy $A>x_0$, and can be obtained by solving the transcendental equation using standard nonlinear root-finding techniques. A graphical examination of Eq.~(\ref{eq:18}) confirms the presence of a single, stable non-trivial root when \( 1 \le \tilde{q} < 2 \), as illustrated in Fig.~\ref{fig:11}. The non-trivial root of \( \tilde{f}(A) \) corresponds to the steady-state amplitude of the limit cycle in the physical system. As $\tilde{q}$ approaches 2, the energy input from the discontinuous forcing significantly outweighs the modal dissipation. For $\tilde{q} \ge 2$, the function $\tilde{f}(A)$ no longer intersects the zero-axis for $A > x_0$, resulting in the unbounded growth mentioned in Section \ref{section_3}. A summary of the stability criteria is shown for this SDOF system in Table \ref{tab4}.
\begin{table}[!ht]
\caption{Stability criteria for the SDOF system given by Eq.~(\ref{eq:16}), whose envelope is governed by Eq.~(\ref{eq:18})}
\label{tab4}
\centering
\renewcommand{\arraystretch}{1}
\begin{tabular}{@{}lll@{}}
\toprule
\textbf{Solution} & \textbf{Existence} \ & \textbf{Criteria for stability} \\
\midrule
$A^* = 0$ & Always & $ \tilde{q} < 1$ \\
$A^* \neq 0 $ & $\strut 1 < \tilde{q} < 2$   & $\strut 1 < \tilde{q} < 2$  \\
\bottomrule
\end{tabular}
\end{table}

Throughout the MDOF analysis, we restrict the parameter range to $\tilde{q}_i < 2$ to ensure the existence of stable limit cycles.

\bibliographystyle{asmejour}   %% .bst file that follows ASME journal format. Do not change.

\bibliography{asmejour_AC_RG.bib} 

%%%%%%%%%%%%%%%%%%%%%%%%%%%%%%%%%%%%%%%%%%%%%%%%%%%%%%%%%%%%%%%%%%%%%%

%% To omit final list of figures and tables, use the class option [nolists]

\end{document}